\documentclass[epj]{svjour}

\usepackage{dcolumn}
\usepackage{bm}
\usepackage{footnote}
\makesavenoteenv{table*}
\makesavenoteenv{tabular}

\usepackage{graphicx}

\usepackage[square,numbers,sort,compress]{natbib}

\usepackage{amsmath}
\usepackage{amssymb}
\usepackage{gensymb}

\usepackage{hyperref}

\begin{document}


\title{Interpretations of Elastic Electron Scattering}

\author{T.W.~Donnelly \and D.K.~Hasell \and R.G.~Milner}
\institute{Center for Theoretical Physics, Laboratory for Nuclear
  Science, and Department of Physics\\Massachusetts Institute of
  Technology, Cambridge, MA 02139}
\mail{R.G.~Milner \email{milner@mit.edu}}

\date{\today}

\abstract{ Elastic scattering of relativistic electrons from the
  nucleon yields Lorentz invariant form factors that describe the
  fundamental distribution of charge and magnetism.  The spatial
  dependence of the nucleon's charge and magnetism is typically
  interpreted in the Breit reference frame which is related by a
  Lorentz boost from the laboratory frame, where the nucleon is at
  rest.  We construct a toy model to estimate how the charge and
  magnetic radii of the nucleon are modified between the Breit and
  lab. frames.  This has implications for the ratio of the proton
  electric to magnetic elastic form factors as a function of momentum
  transfer as well as for determinations of the proton charge radius.
  Predicted corrections based on the model are provided for the 
  rms charge radii of the deuteron,
  the triton, and the helium isotopes. 
}

\maketitle

Elastic scattering of relativistic electrons from the nucleon and
nuclei yields their fundamental distributions of charge and magnetism
in terms of Lorentz invariant form factors~\citep{Hofstadter:1961aa}.
In the case of the nucleon, modern CW electron accelerators using
polarization observables have enabled the determination of the four
elastic proton and neutron electric and magnetic form factors with
precision over a large range of momentum
transfer~\citep{Arrington:2011kb}. This increased precision has yielded
a number of surprises.  The ratio of electric to magnetic form factors
determined using polarization transfer disagrees with the previous
cross section data --- the former is seen to decrease dramatically
with increasing momentum transfer~\citep{Punjabi:2014tna}.  In fact,
while viewed as a surprise, this behavior was always expected in
models based on vector meson dominance (see the discussions in
\citep{Crawford:2010gv,Lomon:2006sf} and references
therein). Furthermore, the charge radius of the proton as determined
from precision elastic electron-proton
scattering~\citep{Bernauer:2010wm} disagrees with a high precision
determination obtained from the Lamb shift in muonic
hydrogen~\citep{Pohl:2010zza,Antognini:1900ns}.  This latter
discrepancy has sparked considerable
interest~\citep{Pohl:2013yb,Bernauer:2014cwa}.  In this Letter we
explore how the nucleon's charge and magnetism can be affected by the
choice of reference frame.  We note that the boost corrections have
been previously derived for the charge
distribution~\citep{Licht:1970de,DeForest:1984qe,Robson:2013nwa,Giannini:2013bra}:
here we develop a more general model that we apply to both charge and
magnetic form factors at the largest momentum transfers where they
have been measured.

Consider relativistic elastic electron scattering from the nucleon.
In single photon exchange approximation, the unpolarized elastic eN
scattering cross section in the lab. system (nucleon at rest) can be
written
\begin{equation}
\frac{d\sigma }{d\Omega }=\sigma _{M}f_{rec}^{-1}\cdot F^{2},  \label{dsigma}
\end{equation}%
where the Mott cross section is%
\begin{equation}
\sigma _{M}=\left[ \frac{\alpha \cos \theta /2}{2 E_e \sin ^{2}\theta /2}%
\right] ^{2}  \label{sigmamott}
\end{equation}%
and the recoil factor is $f_{rec}=E_e / E_e^{\prime }$.  Here the
incident electron has energy $E_e $, the scattered electron has energy
$E_e^{\prime }$ and the scattering is through angle $\theta $; also
$\alpha $ is the fine-structure constant. In all expressions, the
extreme relativistic limit is taken, namely the electron mass is
ignored with respect to its energy.

The square of the form factor, $F^2$, can be expressed in terms of the
electric and magnetic Sachs form factors $G_{E}^{p,n}$ and
$G_{M}^{p,n}$ of the proton and neutron, respectively. The
normalizations at $Q^2 = 0$ are
$G_{E}^{p}(0)=G_{M}^{p}(0)/\mu_{p}=G_{M}^{n}(0)/\mu _{n}=1$ (the
neutron charge form factor is irregular, but may be fixed with respect
to the proton case).  As usual, one has%
\begin{eqnarray}
W_{1} &=&\tau G_{M}^{2}  \label{eq5} \\
W_{2} &=&\frac{1}{1+\tau }\left[ G_{E}^{2}+\tau G_{M}^{2}\right] ,
\label{eq6}
\end{eqnarray}%
where $\tau =|Q^{2}| / 4m_{N}^{2}$ with $m_{N}$ the nucleon
mass. Using these the square of the elastic form factor can be written
in several equivalent forms:%
\begin{eqnarray}
F^{2} &=&W_{2}+2W_{1}\tan ^{2}\theta /2  \label{eq11} \\
&=&\frac{1}{(1+\tau ) \epsilon} \left[ \epsilon G_{E}^{2}+\tau G_{M}^{2}%
\right]  \label{eq12} \\
&=&v_{L}W_{L}+v_{T}W_{T},  \label{eq13}
\end{eqnarray}%
where the relative flux of longitudinally polarized virtual photons
$\epsilon$ is given by%
\begin{equation}
\epsilon^{-1}=1+2(1+\tau )\tan ^{2}\theta /2  \ . \label{eq14}
\end{equation}%
Using the fact that for elastic eN scattering one has%
\begin{equation}
\rho \equiv \left\vert \frac{Q^{2}}{q^{2}}\right\vert =\frac{1}{1+\tau }
\label{eq15}
\end{equation}%
we obtain%
\begin{eqnarray}
v_{L} &=&\rho ^{2}=\frac{1}{(1+\tau )^{2}}  \label{eq16} \\
v_{T} &=&\frac{1}{2}\rho +\tan ^{2}\theta /2=\frac{1}{2(1+\tau )}\epsilon%
^{-1}.  \label{eq17}
\end{eqnarray}%
The longitudinal and transverse responses are related to the $W$s or
to the Sachs form factors via%
\begin{eqnarray}
W_{L} &=&(1+\tau )G_{E}^{2}  \label{eq18} \\
W_{T} &=&2\tau G_{M}^{2}.  \label{eq19}
\end{eqnarray}

One definition of a ``proton charge radius''\ --- the one commonly
used in analyses of elastic electron-proton scattering --- is obtained
through the derivative of $G_{E}^{p}(Q^{2})$ with respect to the
invariant 4-momentum transfer, $Q^{2}$, namely via%
\begin{eqnarray}
\left[ -6\frac{dG_{E}^{p}(Q^{2})}{dQ^{2}}\right]_{Q^2 = 0}\equiv \left( r^{\mathrm{mom}}_{E,p} \right) ^{2}. \label{eqadd1}
\end{eqnarray}%
Since both $G_{E}^{p}(Q^{2}) $ and $Q^{2}$ are invariants, the
so-defined quantity with dimensions of length is also. \ We denote
this type of radius, $r^{\mathrm{mom}}_{E,p}$, as the momentum-space
proton charge radius. It is a convenient quantity to employ when
inter-comparing data on elastic $ep$ scattering. What it is not is the
RMS charge radius of the proton. In principle the latter would be
found by taking the charge distribution of the proton in its rest
frame, weighting by $r^{2}$, integrating and taking the square root to
obtain the coordinate-space charge radius
$r^{\mathrm{coord}}_{E,p}$. In some modeling or approaches such as
lattice QCD one works in coordinate space and proceeds in this
manner. Non-relativistically the two radii are the same, and were the
same logic to be applied to heavy nuclei, even relativistically the
differences would be negligible. However, the nucleon (and to a lesser
extent few-body nuclei) is exceptional: the difference between the two
definitions is larger than the present uncertainties in the data.
Purists would say that only the momentum-space quantity should be used
when discussing elastic $ep$ scattering, since this is intrinsically a
process where different frames are involved. In the lab frame, for
instance, the proton starts at rest, but must recoil after absorbing
the momentum transferred via the exchanged virtual photon; and
choosing the Breit frame does not make the problem go away, for then
the proton must be moving in both the initial and final states (see
below).

Were only electron scattering to be considered, it is hard to argue
with this point of view, and it is largely our non-relativistic bias
that motivates us to attempt to define a quantity that we hope is some
sort of radius. Unfortunately, typical modeling is not done in a fully
covariant way, but rather is done by attempting to provide a picture
in a specific frame (for instance, through some approach such as
lattice QCD or employing models such as the bag model) and then
Fourier transforming to obtain a form factor. For any model that is
not boostable this procedure incurs the same difficulty as the above
for the two types of radius. Furthermore, as discussed briefly later
where electronic and muonic atomic hydrogen are mentioned, there are
other measurements involving the concept of a proton charge radius
that one wants to use to explore whether or not they yield radius
values that are compatible with those found via electron scattering.
An important question is: which type of radius do these other
measurements probe?

Nevertheless, let us not take the point of view that only the
invariant momentum-space radius defined above is relevant and at least
attempt to evaluate the potential uncertainty in employing a specific
definition of the radius (later generalized to all four radii for the
nucleon) with consequence for comparisons of electron scattering
determinations with those from the Lamb shifts in electronic and
muonic hydrogen. We start by summarizing some of the basics in the
Breit frame.

The Breit frame is defined by zero energy transfer of the
virtual photon and reversal of the 3-momentum of the target between
the initial and final states.  Thus, one has $\omega _{B} =0$ and
$q_{B} =\sqrt{|Q^{2}|}$.  Since in that frame (with $z$-axis along the
momentum transfer vector) one has the nucleon entering with 3-momentum
$-p_{B}$ and leaving with 3-momentum $+p_{B}$, one has
$p_{B}=q_{B}/2$.  Thus, the relativistic $\gamma $-factor relative to
the lab. frame for the nucleon in that frame is%
\begin{equation}
\gamma =\sqrt{1+\tau }.  \label{eq23}
\end{equation}
We note that, no matter what reference frame is adopted, the nucleon
before and after the scattering must have different momenta. In the
lab. frame the nucleon begins at rest, but must recoil with the full
momentum in the final state. In the Breit frame (the ``brick-wall''
frame as discussed above) it must enter with momentum $-q_{B}/2$ and
leave with momentum $+q_{B}/2$. In any case, Lorentz boosts are
involved and accordingly it is reasonable to expect some consequences
from these such as Lorentz contractions and time dilations. For
instance, one might expect a moving nucleon to be flattened (as nuclei
are in relativistic heavy-ion collisions) with corresponding
modification of its ``charge distribution''. In a covariant
description in momentum space the spinors and operators are typically
handled appropriately; however, if one tries to interpret the Fourier
transforms, problems arise and concepts such as the charge and spin
distributions are not simply these Fourier transforms, as to a large
extent they are for heavier systems such as nuclei. One might just
work in momentum space and never confront these issues, although there
is strong sentiment that some interpretation should be attempted in
coordinate space and in the following we try to do this with a ``toy
model'' that, while certainly not a fundamental one, is motivated by
our intuition. Note that scattering involves these problems of
reference frames and boosts, whereas the Lamb shift does not have the
same issue for there one has an electron or muon bound to a proton
with a single frame to consider. Of course, these atomic systems have
their own issues (relativistic corrections, recoil effects, {\it
  etc.})

We now attempt to formulate a non-relativistic version of eN
scattering. Assume that one puts a single nucleon into the lowest
level in a very deep harmonic oscillator (HO) potential to avoid any
recoil problem. In effect, in this ``toy model" one can imagine there
being a very strong trap to hold the nucleon essentially at rest;
indeed, a nucleon in a heavy nucleus is confined and not allowed to
carry the full recoil momentum when electrons are scattered from it,
providing an example where this situation actually occurs. Then using,
for instance, the tables of \citep{Donnelly:1979ezn} or the review
article of \citep{DeForest:1966ycn}, one can compute the multipole
matrix elements of the C0 (Coulomb monopole) and M1 (magnetic dipole)
elastic scattering operators.  As discussed in the standard literature
for electron scattering from nuclei, and those papers in particular,
the differential cross section may be written as in eq.~\ref{dsigma}
and~\ref{sigmamott} where, using the non-relativistic limit for the
current operators together with $1s_{1/2}$ harmonic oscillator wave
functions, one obtains the following:%
\begin{eqnarray}
 \left\langle 1s_{1/2}\left\Vert M_{0}^{Coul} \right\Vert
1s_{1/2}\right\rangle  &=& F_{1}\left\langle 1s_{1/2}\left\Vert
M_{0}\right\Vert 1s_{1/2}\right\rangle  \label{eq24} \\
\left\langle 1s_{1/2}\left\Vert  iT_{1}^{mag} \right\Vert
1s_{1/2}\right\rangle  &=& \frac{q}{m_{N}} [ F_{1}\left\langle
1s_{1/2}\left\Vert \Delta _{1}(q\mathbf{x})\right\Vert 1s_{1/2}\right\rangle \nonumber \\
-\frac{1}{2}\left( F_{1}+F_{2}\right) &\cdot& \left\langle 1s_{1/2}\left\Vert \Sigma
_{1}^{\prime }(q\mathbf{x})\right\Vert 1s_{1/2}\right\rangle ] \ . 
\label{eq25}
\end{eqnarray}
Here, following \citep{DeForest:1966ycn}, we use the Dirac and Pauli
single-nucleon form factors, as is conventional for non-relativistic
treatments of electron scattering from nuclei. Either working directly
with the harmonic oscillator wave functions and the explicit forms for
the current multipole operators (see \citep{DeForest:1966ycn}) or (which is
easier) using the tables of \citep{Donnelly:1979ezn} one finds for the three
required reduced matrix elements
\begin{eqnarray}
\sqrt{4\pi }\left\langle 1s_{1/2}\left\Vert M_{0}(q\mathbf{x})\right\Vert
1s_{1/2}\right\rangle  &=&\sqrt{2}e^{-y} \\
\sqrt{4\pi }\left\langle 1s_{1/2}\left\Vert \Delta _{1}(q\mathbf{x}%
)\right\Vert 1s_{1/2}\right\rangle  &=&0 \\
\sqrt{4\pi }\left\langle 1s_{1/2}\left\Vert \Sigma _{1}^{\prime }(q\mathbf{x}%
)\right\Vert 1s_{1/2}\right\rangle  &=&2e^{-y}.
\end{eqnarray}
Each is proportional to $\exp (-y)$ where $y=(bq/2)^{2}$ with $b$ the
HO parameter. However, one must multiply by the center-of-mass
correction which in the non-relativistic HO shell model can be
computed: it is a multiplicative factor of
$f_{cm}=\exp (+y/A)=\exp (+y)$ for $A=1$ and cancels the above factor,
leaving only the remaining factors obtained by using the above-cited
tables. Performing the detailed developments one can obtain the form
factors in the HO shell model and, comparing the covariant expressions
in eq.~\ref{eq18} and~\ref{eq19} with these results, one can establish a
correspondence between the non-relativistic so-defined quantities and
their relativistic counterparts%
\begin{eqnarray}
{\rm Non-relativistic}&  &{\rm Relativistic}\nonumber\\
& &\nonumber\\
F_{L}^{2} = F_1^2 &\quad\leftrightarrow\quad& (1+\tau )G_{E}^{2} \\
F_{T}^{2} =\left[ \frac{q}{2 m_N} \right]^2 (F_1 + F_2)^2  &\quad\leftrightarrow\quad& 2\tau G_{M}^{2}
\end{eqnarray}%
where $F_1$ and $F_2$ are the Dirac and Pauli form factors of the
nucleon. Note that for the transverse form factor the left-hand
expression involves the square of the 3-momentum transfer, whereas the
right-hand side involves the square of the 4-momentum
transfer. Equivalently, to ``de-relativize'' the covariant expressions
one should define the following non-relativistic expressions:%
\begin{eqnarray}
G_{E}^{nr} &\equiv &\sqrt{1+\tau }G_{E}^{rel}  \label{eq27} \\
G_{M}^{nr} &\equiv &\frac{1}{\sqrt{1+\tau }}G_{M}^{rel},  \label{eq28}
\end{eqnarray}%
namely, modified in the opposite ways by the factor $\gamma $ (see
eq.~\ref{eq23}). The motivation is to go from the covariant case where
boosts (Lorentz contractions and time dilations) must occur, since, in
whatever frame one chooses, the nucleon must have non-zero momentum at
some point in the scattering process, to a frame that is more akin to
our non-relativistic intuition.

\begin{figure*}[!ht]
\centering
\includegraphics[viewport = 0 0 519 354, width=0.9\textwidth,
clip]{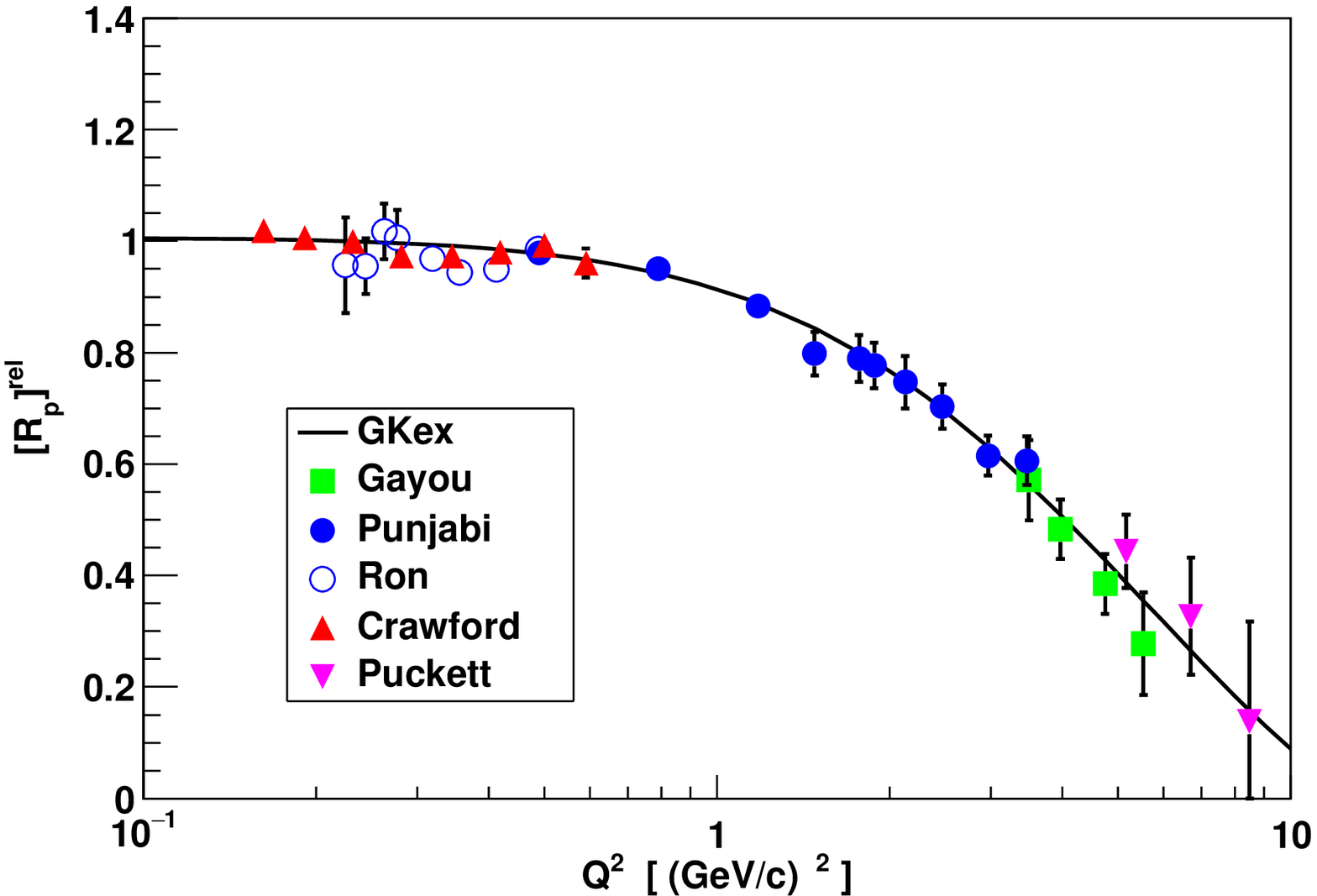}
\caption{(Color online) $\left[ R_p\right]^{rel}$ vs. $Q^2$ for the data
  in~\citep{Gayou:2001qd,Punjabi:2005wq,Ron:2007vr,Crawford:2006rz,Puckett:2010ac}
  together with the VMD-based curve~\citep{Crawford:2010gv,Lomon:2006sf}.}
\label{rvsQ2}
\end{figure*}

\begin{figure*}[!ht]
\centering
\includegraphics[viewport = 0 0 552 354, width=0.9\textwidth, clip]{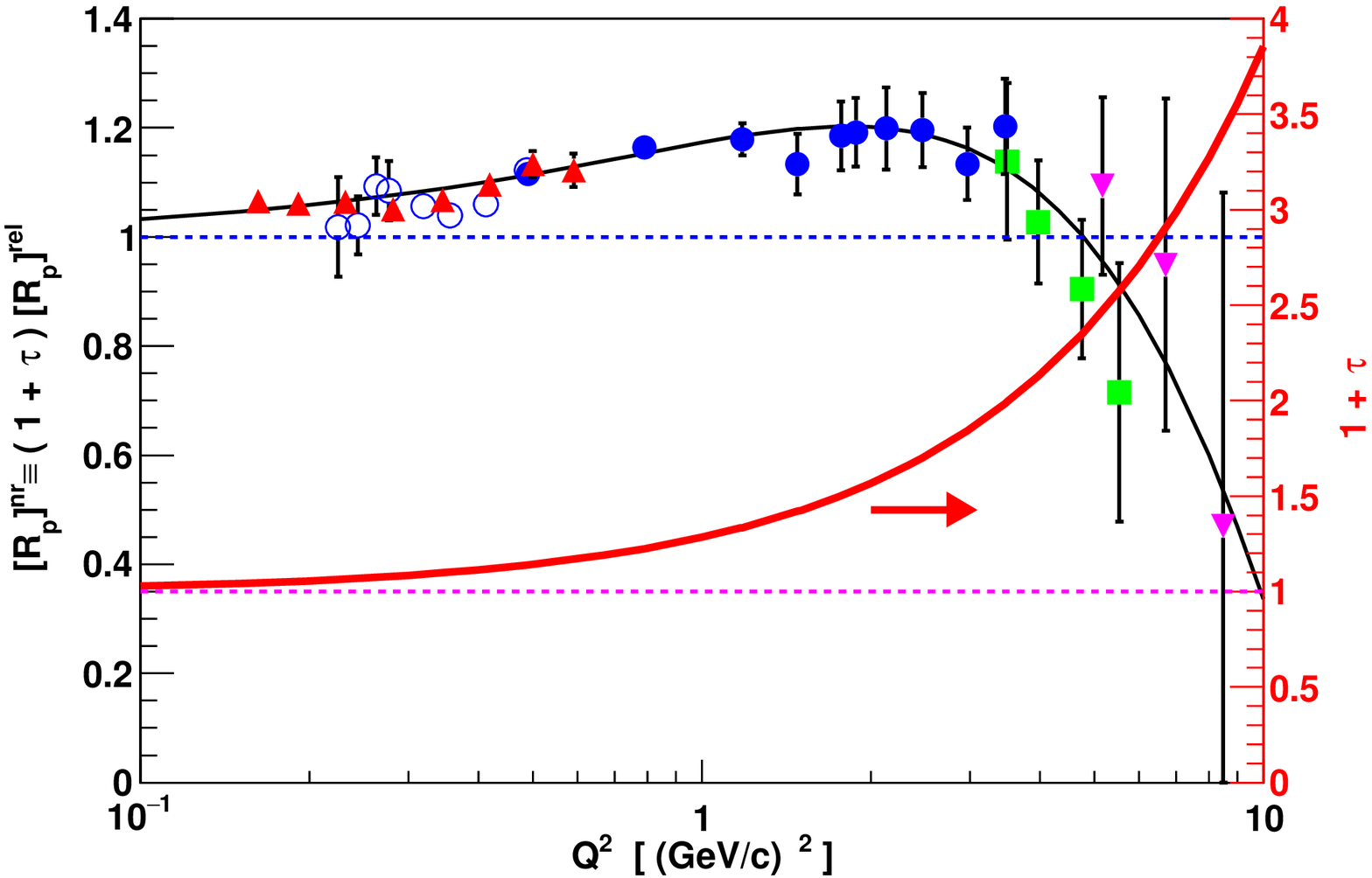}
\caption{(Color online) Left panel: $\left[ R_p\right]^{nr} \equiv (1+\tau)\left[ R_p\right]^{rel}$
  vs. $Q^2$ with data and GKex curve as in Fig.~\ref{rvsQ2}, together
  with $1+\tau$ curve referenced to the right-hand axis.}
\label{taurvsQ2}
\end{figure*}

The usual form factor ratio is defined as follows%
\begin{equation}
\left[ R_{p} \right]^{rel}\equiv \frac{G_{E}^{p}}{G_{M}^{p}/\mu _{p}}  \label{eq29}
\end{equation}%
and is shown in Fig.~\ref{rvsQ2}; see \citep{Crawford:2010gv} for
references to the data and to the so-called GKex vector meson based
model \citep{Lomon:2006sf} shown as a solid line in the figure. The
``de-relativized'' ratio is then immediately given by%
\begin{equation}
\left[ R_{p}\right] ^{nr}=(1+\tau )\left[ R_{p}\right] ^{rel}  \label{eq30}
\end{equation}%
and is shown in Fig.~\ref{taurvsQ2}. Note that this has the
boost factor squared (going as $1+\tau$ and shown in the right panel
as a red line) rather than just linearly as in the individual form
factors. Clearly this introduces large modifications at high momentum
transfers. Indeed, the ``de-relativized'' results are relatively flat
as functions of $Q^2$ and differ from unity by less than roughly 20\%.


Next, one might attempt to ``de-relativize" the definition of the RMS
radii using the same procedures. The usual definition is obtained by
expanding the $j_{0}$ spherical Bessel function for low momentum
transfer to obtain%
\begin{equation}
r_{E}^{rel}\equiv \sqrt{\frac{3}{2m_{N}^{2}}\left\vert \left( \frac{d}{d\tau
}G_{E}\right) _{\tau \rightarrow 0}\right\vert };  \label{eq31}
\end{equation}%
this is simply a re-writing of eq.~\ref{eqadd1}. That is, we
identify $r_E^{rel}$ with what was called $r_E^{mom}$ at the beginning
of the paper. We then proceed to ``de-relativize" to obtain a quantity
that might be more intuitive, based on our toy model. For this one
must include the $\sqrt{1+\tau }$ factor in eq.~\ref{eq27},
obtaining at small momentum transfer
\begin{eqnarray}
G_{E}^{nr} &=&\left( 1+\frac{1}{2}\tau +\cdots \right) \nonumber\\
 &&\quad\cdot \left( 1-\frac{2}{3}\tau m_{N}^{2}\left[ r_{E}^{rel}\right] ^{2}+\cdots \right)\\&=&
1-\left[ \frac{2}{3}m_{N}^{2}\left[ r_{E}^{rel}\right] ^{2}-\frac{1}{2} \right] \tau +\cdots   \\
&\equiv&1-\frac{2}{3}\tau m_{N}^{2}\left[ r_{E}^{nr}\right]^{2}+\cdots \label{eq34}
\end{eqnarray}
and leading to the relationship
\begin{equation}
r_{E}^{nr}=\sqrt{\left[ r_{E}^{rel}\right] ^{2}-\Delta },  \label{eq35}
\end{equation}%
where one has%
\begin{eqnarray}
\Delta _{p} &\equiv &\frac{3}{4m_{p}^{2}}=0.0332\mathrm{\ fm}^{2}
\label{eq35a} \\
\Delta _{n} &\equiv &\frac{3}{4m_{n}^{2}}=0.0331\mathrm{\ fm}^{2}
\label{eq35b}
\end{eqnarray}%
for protons and neutrons, respectively. The same arguments for the
magnetic form factor where the required boost factor is now
$1/\sqrt{1+\tau}$ (see eq.~\ref{eq28}) leads to the expression%
\begin{equation}
r_{M}^{nr}=\sqrt{\left[ r_{M}^{rel}\right] ^{2}+\Delta }.  \label{eq36}
\end{equation}%
We note that the same result for the proton charge form factor was
obtained in \citep{Giannini:2013bra}, arguing from a very different point of view:
see also~\cite{Licht:1970de}, on which that work is based.

It is important to understand that these differences between
relativistic and non-relativistic radii do not go away if electron
scattering data are obtained at ever smaller values of the momentum
transfer. As the above expressions clearly show, the relativistic
boost factor arising from $(1+\tau)^{\pm 1/2}$ deviates from unity at
order $Q^2$; however, that is the order needed to extract the charge
or magnetic radii. In other words, being locked together at the same
order when expanding in powers of $Q^2$ the effects can never be
separated, no matter how small the momentum transfer becomes.

Specifically, using the Bernauer value for the proton rms charge radius~\cite{Bernauer:2010wm},
and the PDG values~\citep{Agashe:2014kda} for the proton magnetic, neutron charge,
and neutron magnetic rms radii, one has the following:
\begin{eqnarray}
r_{E,p}^{nr} &=& \sqrt{(0.879)^{2}-0.0332}\nonumber\\
&=&0.860\pm 0.008~{\rm  fm}\\
r_{M,p}^{nr} &=&\sqrt{(0.777)^{2}+0.0332}\nonumber\\ 
&=& 0.7981\pm 0.013\pm 0.010~{\rm fm}\\ 
r_{M,n}^{nr} &=&\sqrt{(0.862)^{2}+0.0331}\nonumber\\
&=&0.8810^{+0.009}_{-0.008}~{\rm fm}\\
\left[ r_{E,n}^{nr}\right] ^{2} &=&-0.1161-0.0331\nonumber\\
&=&-0.1492\pm 0.0022~{\rm fm}^{2} \ .
\end{eqnarray}
Here the uncertainties are taken from the Bernauer work~\cite{Bernauer:2010wm} and from the PDG
compilation~\citep{Agashe:2014kda}, respectively. Note that the electric result for the neutron
is traditionally expressed as the square of the radius, which is
negative.

What has been obtained above from electron scattering through the
simple toy model are radii that start from the quantities measured in
electron scattering ($r^{mom} \equiv r^{rel}$) to quantities that are
more like $r^{\mathrm{coord}}$. The toy model employed in this work
thus leads us to make the identification
$r^{coord} \leftrightarrow r^{nr}$.  Namely, when coordinate-space
radii are wanted, rather than the momentum-space ones measured in
electron scattering, the toy model motivates taking into account the
corrections discussed above. While the model is not ``fundamental'',
at least the fact that the relativistic and non-relativistic
quantities so-defined are different should be cause for some concern
that the concept of a charge or magnetic radius is not totally robust.

Let us take this one step further and bring in the values of the
\textquotedblleft proton charge radius\textquotedblright\ determined
via the Lamb shifts in electronic and muonic hydrogen which are known
to have some charge-distribution dependence because the proton is not
a point particle, but has a finite extent. Note that these atomic
systems are described in coordinate space and not, as for electron
scattering, in momentum space. In the course of developing the
formalism for studies of atomic hydrogen it is natural to invoke the
Fourier transform of the proton charge distribution to inter-relate
what is desired in this case with what is measured in electron
scattering; however, this is the essence of the issue. Namely, it
would appear to be more natural that the Lamb shift problem involves
the coordinate-space radius $r^{\mathrm{coord}}_{E,p}$ and not its
momentum-space analog, \textit{i.e.,} to use $r_{E,p}^{nr}$, not its
relativistic partner. One might be tempted to use the fact that the
effective value of $Q r_{E,p}$ is very small for these atomic systems
($\sim 10^{-5}$); however, the above argument on the ``locking'' of
the boost factor with the radius shows that the smallness of this
product is not sufficient to make relativistic and non-relativistic
radii effectively the same. It is not the scale of momentum transfer
that is critical (as long as it is small enough to allow only terms of
quadratic order to be considered), but the fact that a scattering
process and measurements of an atomic system are more naturally
studied in momentum space and coordinate space, respectively. We
assume that analysis of the Lamb shift entails using the
coordinate-space version of the proton charge radius which, in our toy
model, means $r_{E,p}^{nr}$.

\begin{figure*}[!ht]
\centering
\includegraphics[viewport=0 0 512 354, width=0.9\textwidth, clip]{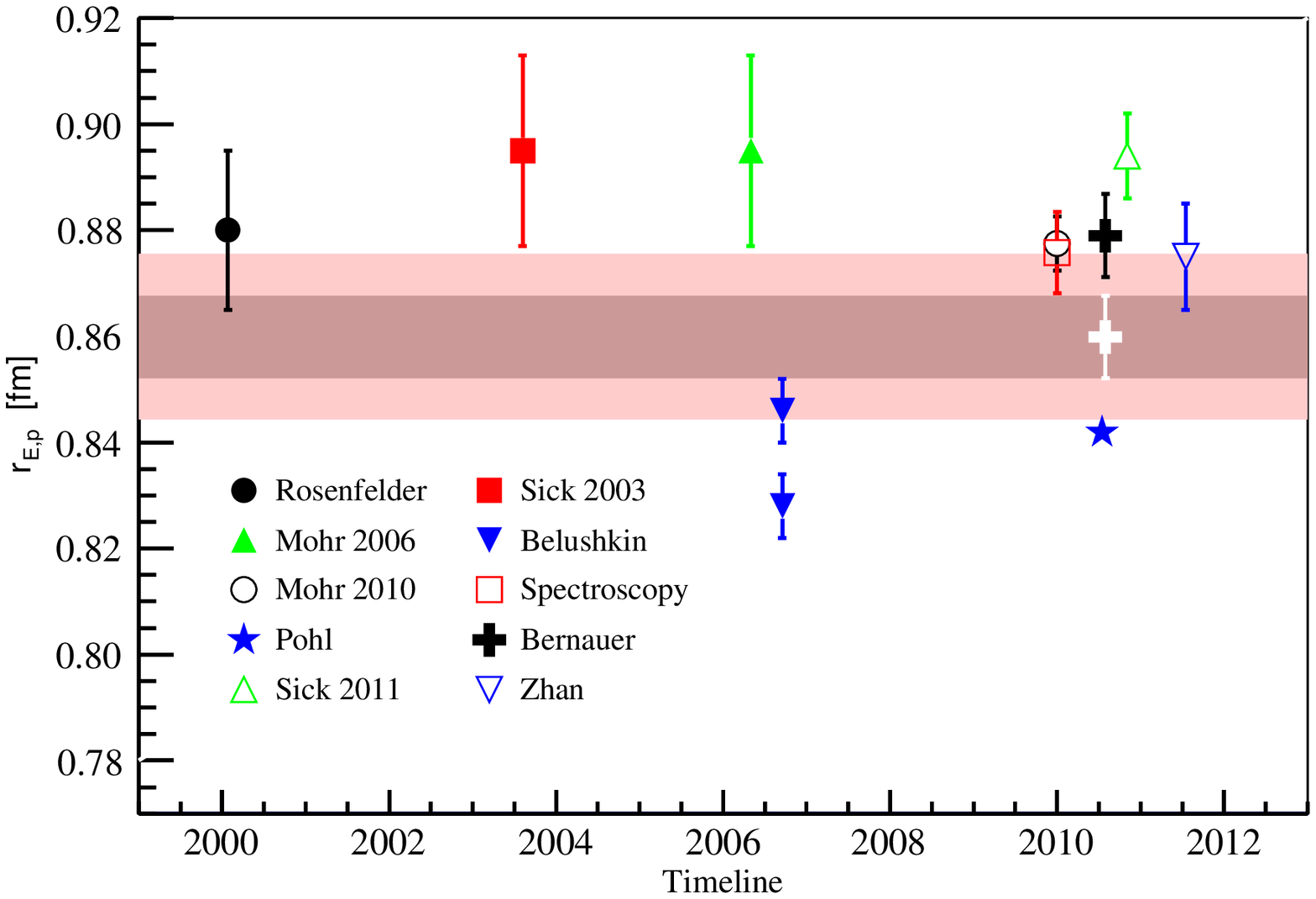}
\caption{Timeline of recent
  determinations~\citep{Rosenfelder:1999cd,Sick:2003gm,Mohr:2008fa,
  Belushkin:2006qa,Mohr:2012tt,Pohl:2013yb,Bernauer:2010wm,
  Sick:2011zz,Zhan:2011ji} of
  the proton charge radius $r_{E,p}$.  For earlier experimental results see~\cite{Pohl:2010zza}.
  The Bernauer value for the
  proton charge radius, namely $r^{mom}_{E,p} = r^{rel}_{E,p}$, is
  indicated with a black cross lying at 0.879, while its corrected
  value $r^{coord}_{E,p} = r^{nr}_{E,p}$ is indicated by the white
  cross at 0.860. The shaded bands show the 1$\sigma$ and 2$\sigma$
  uncertainties about the latter and may then be compared with the
  Lamb shift value indicated by a star and lying at about 0.84. }
\label{timeline}
\end{figure*}

The proton charge radius discrepancy has arisen from the different
values resulting from a precise determination using the Lamb shift in
muonic hydrogen (0.84087 $\pm$ 0.00026 $\pm$ 0.00029)~\citep{Antognini:1900ns}
disagreeing with the CODATA 2010 value (0.8775 $\pm$
0.0051)~\citep{Mohr:2012tt}, largely determined by the most precise value
resulting from elastic electron proton scattering~\citep{Bernauer:2010wm}.  This
amounts to more than 4\% difference, whereas the stated total uncertainty in
the Bernauer electron scattering result is quoted as 0.9\%. On the other hand,
the correction resulting from the boost between Breit and lab. frames
as calculated in our toy model decreases the electron scattering
determination of $r_{E,p}$ towards the muonic hydrogen value.  The
resulting discrepancy in the different determinations of $r_{E,p}$ is
cut in half using the corrected value and now differs from the muonic
Lamb shift value by only about 2\%. Further, the corrected value here
for the proton charge radius is not inconsistent with the value
determined using the hydrogen atom, within experimental uncertainty.
Fig.~\ref{timeline}
shows a selection of determinations of the proton rms charge radius since the year 2000
including all the work cited here.  In particular, it shows the electron scattering determination
from  Bernauer and our corrected value with both 1$\sigma$ and 2$\sigma$ uncertainties.
If one accepts the ideas we have developed in this paper, then one is left with only a small amount
to explain and, the remaining discrepancy being commensurate with the
size of $\alpha$, might lead one to question the other model
dependences in the problem (experimental systematic uncertainties,
radiative corrections, {\it etc.})

The $Q^2$-independent recoil correction to the rms charge radius
derived from elastic electron scattering can also be evaluated for the
deuteron, the triton, and the helium isotopes.  For heavier nuclei,
the correction factor above is replaced as follows:
\begin{equation}
\Delta = \frac{3}{4m_{N}^{2}}\rightarrow \Delta\cdot \left( \frac{m_{N}%
}{m_{A}}\right) ^{2},  \label{eq37}
\end{equation}%
where $m_{A}$ is the mass of that heavier system.
In Table~\ref{tableRp}, 
\begin{table*}
\centering
\begin{tabular}{|r|c|c|c|}
\hline 
Nucleus & $r_{E}$ & $r_{E}$ & $r_{E}$ \\ 
              & e-scattering & e-scattering corrected &   muonic atom \\ 
               & (fm)                 & (fm) & (fm) \\ \hline
$^1$H & 0.879 $\pm$ 0.008~\cite{Bernauer:2010wm} &  0.860 $\pm$ 0.008 & 0.84087 $\pm$ 0.00039 \\
$^2$H & 2.13 $\pm$ 0.01~\cite{Sick:2008aa} &  2.12 $\pm$ 0.01 &
       2.12562 $\pm$ 0.00078~\cite{Krauth:2016aa} \\
$^3$H  & 1.755 $\pm$ 0.087~\cite{Sick:2008aa} &  1.751 $\pm$ 0.087&  \\
$^3$He & 1.959 $\pm$ 0.034~\cite{Sick:2008aa} & 1.955 $\pm$  0.034 & \\
$^4$He  &  1.680 $\pm$ 0.005~\cite{Sick:2008aa} & 1.678 $\pm$ 0.005 &  \\
\hline
\end{tabular}
\caption{Comparison of charge radii measured by electron scattering
  from various light nuclei and the corrected values following the
  procedure proposed herein.  The muonic determinations for the proton and the deuteron (preliminary
  value - see~\cite{Krauth:2016aa})
  are also given.  We note that high precision values for the helium isotopes 
  will be forthcoming from muonic atom experiments.}
\label{tableRp}
\end{table*}
we summarize the results based on our model.  We note that the
corrected deuteron rms charge radius from electron scattering is in
excellent agreement with the recent high precision value obtained from
measurements of the Lamb shift in muonic deuterium.  We provide
predictions for the rms charge radii of the helium isotopes, which are
being determined to high precision in ongoing experiments that employ
measurement of the Lamb shift in muonic atoms.  New high precision
measurements of the Lamb shift in electronic hydrogen are in progress
and results are expected soon.


We have also considered how the conclusions here can be validated by
experiment.  We point out that the correction we derive cannot be
separated by going to lower $Q^2$ in electron scattering experiments
nor by comparison of elastic electron and muon scattering on the proton.
However, based on the estimates made here, a precision comparison of
electron scattering from the proton with electron scattering from the
deuteron at low $Q^2$ should deviate at the level of about 1.5\%, since the
corrections we derive differ at this level.  For highest precision,
such measurements should be carried out using the same apparatus and
systematics must be minimized.  We note that internal radiative
corrections, which arise mainly from the incident and scattered
electrons, should be quite similar for proton and deuteron targets.

In summary, we argue that corrections between the Breit and lab.
frames are important in interpreting the form factors of the nucleon
as determined in relativistic electron scattering.  We have
constructed a toy model to estimate these corrections.  In this toy
model, the observed significant decrease of the ratio of the proton
elastic form factors as a function of $Q^2$ is understood as a
predominantly relativistic effect.  Furthermore, in this toy model,
the proton charge radius as determined in electron scattering has a
correction that reduces its value towards that resulting from the
precision determination using the Lamb shift in muonic hydrogen. The
toy model provides predictions for the rms charge radii of the deuteron, triton,
and helium isotopes. 

The ideas presented here should not be viewed as providing a detailed
model of how boost effects might enter in electron scattering from the
nucleon, but as a warning that they could play a role in electron
scattering, but not in determinations of the Lamb shift. Much more
satisfying would be to have a model of the nucleon that could be
boosted, although it is not obvious that such a model exists at
present. For instance, in models where relativistic quarks are
confined via a ``bag" one must confront the problem of how to boost
the latter in going between the frames that inevitably enter in
electron scattering. One possible, but different, related case that
could be studied to test the ideas is that of relativistic (covariant,
boostable) modeling of the deuteron~\citep {VanOrden:2016aa}.  There, on the one
hand, one could directly compute the elastic form factor, while on the
other hand one could compute the ground-state charge distribution and
then Fourier transform it to momentum space.  Upon comparing the two
results it is likely that differences will be found that relate to the
boost issues raised in the present study.

\begin{acknowledgement}
We thank Jan Bernauer for valuable discussion. The authors' research
is supported by the Office of Nuclear Physics of the U.S. Department
of Energy under grant Contract Numbers DE-SC0011090 and
DE-FG02-94ER40818.
\end{acknowledgement}

\bibliographystyle{mybibstyle} 
\bibliography{FormFactor-epj}

\begin{thebibliography}{10}
\expandafter\ifx\csname url\endcsname\relax
  \def\url#1{\texttt{#1}}\fi
\expandafter\ifx\csname urlprefix\endcsname\relax\def\urlprefix{URL }\fi
\providecommand{\bibinfo}[2]{#2}
\providecommand{\eprint}[2][]{\url{#2}}

\bibitem{Hofstadter:1961aa}
\bibinfo{author}{Hofstadter, R.}
\newblock \emph{\bibinfo{journal}{Nobel Lecture}}  (\bibinfo{year}{1961}).

\bibitem{Arrington:2011kb}
\bibinfo{author}{Arrington, J.}, \bibinfo{author}{de~Jager, K.} \&
  \bibinfo{author}{Perdrisat, C.~F.}
\newblock \emph{\bibinfo{journal}{J. Phys. Conf. Ser.}}
  \textbf{\bibinfo{volume}{299}}, \bibinfo{pages}{012002}
  (\bibinfo{year}{2011}).

\bibitem{Punjabi:2014tna}
\bibinfo{author}{Punjabi, V.} \& \bibinfo{author}{Perdrisat, C.~F.}
\newblock \emph{\bibinfo{journal}{EPJ Web Conf.}}
  \textbf{\bibinfo{volume}{66}}, \bibinfo{pages}{06019} (\bibinfo{year}{2014}).

\bibitem{Crawford:2010gv}
\bibinfo{author}{Crawford, C.~{\it et al.}.}
\newblock \emph{\bibinfo{journal}{Phys. Rev.}} \textbf{\bibinfo{volume}{C82}},
  \bibinfo{pages}{045211} (\bibinfo{year}{2010}).

\bibitem{Lomon:2006sf}
\bibinfo{author}{Lomon, E.~L.}
\newblock \emph{\bibinfo{journal}{arXiv:nucl-th/0609020v2}}
  (\bibinfo{year}{2006}).

\bibitem{Bernauer:2010wm}
\bibinfo{author}{Bernauer, J. C.~{\it et al.}.}
\newblock \emph{\bibinfo{journal}{Phys. Rev. Lett.}}
  \textbf{\bibinfo{volume}{105}}, \bibinfo{pages}{242001}
  (\bibinfo{year}{2010}).

\bibitem{Pohl:2010zza}
\bibinfo{author}{Pohl, R.~{\it et al.}.}
\newblock \emph{\bibinfo{journal}{Nature}} \textbf{\bibinfo{volume}{466}},
  \bibinfo{pages}{213--216} (\bibinfo{year}{2010}).

\bibitem{Antognini:1900ns}
\bibinfo{author}{Antognini, A.~{\it et al.}.}
\newblock \emph{\bibinfo{journal}{Science}} \textbf{\bibinfo{volume}{339}},
  \bibinfo{pages}{417--420} (\bibinfo{year}{2013}).

\bibitem{Pohl:2013yb}
\bibinfo{author}{Pohl, R.}, \bibinfo{author}{Gilman, R.},
  \bibinfo{author}{Miller, G.~A.} \& \bibinfo{author}{Pachucki, K.}
\newblock \emph{\bibinfo{journal}{Ann. Rev. Nucl. Part. Sci.}}
  \textbf{\bibinfo{volume}{63}}, \bibinfo{pages}{175--204}
  (\bibinfo{year}{2013}).

\bibitem{Bernauer:2014cwa}
\bibinfo{author}{Bernauer, J.~C.} \& \bibinfo{author}{Pohl, R.}
\newblock \emph{\bibinfo{journal}{Sci. Am.}} \textbf{\bibinfo{volume}{310}},
  \bibinfo{pages}{18--25} (\bibinfo{year}{2014}).

\bibitem{Licht:1970de}
\bibinfo{author}{Licht, A.~L.} \& \bibinfo{author}{Pagnamenta, A.}
\newblock \emph{\bibinfo{journal}{Phys. Rev.}} \textbf{\bibinfo{volume}{D2}},
  \bibinfo{pages}{1156--1160} (\bibinfo{year}{1970}).

\bibitem{DeForest:1984qe}
\bibinfo{author}{De~Forest, T.}
\newblock \emph{\bibinfo{journal}{Nucl. Phys.}}
  \textbf{\bibinfo{volume}{A414}}, \bibinfo{pages}{347--358}
  (\bibinfo{year}{1984}).

\bibitem{Robson:2013nwa}
\bibinfo{author}{Robson, D.}
\newblock \emph{\bibinfo{journal}{Int. J. Mod. Phys.}}
  \textbf{\bibinfo{volume}{E23}}, \bibinfo{pages}{1450090}
  (\bibinfo{year}{2015}).

\bibitem{Giannini:2013bra}
\bibinfo{author}{Giannini, M.~M.} \& \bibinfo{author}{Santopinto, E.}
\newblock \emph{\bibinfo{journal}{arXiv:hep-ph/1311.0319}}
  (\bibinfo{year}{2013}).

\bibitem{Donnelly:1979ezn}
\bibinfo{author}{Donnelly, T.~W.} \& \bibinfo{author}{Haxton, W.~C.}
\newblock \emph{\bibinfo{journal}{Atom. Data Nucl. Data Tabl.}}
  \textbf{\bibinfo{volume}{23}}, \bibinfo{pages}{103--176}
  (\bibinfo{year}{1979}).

\bibitem{DeForest:1966ycn}
\bibinfo{author}{De~Forest~Jr, T.} \& \bibinfo{author}{Walecka, J.~D.}
\newblock \emph{\bibinfo{journal}{Adv. Phys.}} \textbf{\bibinfo{volume}{15}},
  \bibinfo{pages}{1--109} (\bibinfo{year}{1966}).

\bibitem{Gayou:2001qd}
\bibinfo{author}{Gayou, O.~{\it et al.}.}
\newblock \emph{\bibinfo{journal}{Phys. Rev. Lett.}}
  \textbf{\bibinfo{volume}{88}}, \bibinfo{pages}{092301}
  (\bibinfo{year}{2002}).

\bibitem{Punjabi:2005wq}
\bibinfo{author}{Punjabi, V.~{\it et al.}.}
\newblock \emph{\bibinfo{journal}{Phys. Rev.}} \textbf{\bibinfo{volume}{C71}},
  \bibinfo{pages}{055202} (\bibinfo{year}{2005}).

\bibitem{Ron:2007vr}
\bibinfo{author}{Ron, G.~{\it et al.}.}
\newblock \emph{\bibinfo{journal}{Phys. Rev. Lett.}}
  \textbf{\bibinfo{volume}{99}}, \bibinfo{pages}{202002}
  (\bibinfo{year}{2007}).

\bibitem{Crawford:2006rz}
\bibinfo{author}{Crawford, C. B.~{\it et al.}.}
\newblock \emph{\bibinfo{journal}{Phys. Rev. Lett.}}
  \textbf{\bibinfo{volume}{98}}, \bibinfo{pages}{052301}
  (\bibinfo{year}{2007}).

\bibitem{Puckett:2010ac}
\bibinfo{author}{Puckett, A. J. R.~{\it et al.}.}
\newblock \emph{\bibinfo{journal}{Phys. Rev. Lett.}}
  \textbf{\bibinfo{volume}{104}}, \bibinfo{pages}{242301}
  (\bibinfo{year}{2010}).

\bibitem{Agashe:2014kda}
\bibinfo{author}{Olive, K. A.~{\it et al.}.}
\newblock \emph{\bibinfo{journal}{Chin. Phys.}} \textbf{\bibinfo{volume}{C38}},
  \bibinfo{pages}{090001} (\bibinfo{year}{2014}).

\bibitem{Mohr:2012tt}
\bibinfo{author}{Mohr, P.~J.}, \bibinfo{author}{Taylor, B.~N.} \&
  \bibinfo{author}{Newell, D.~B.}
\newblock \emph{\bibinfo{journal}{Rev. Mod. Phys.}}
  \textbf{\bibinfo{volume}{84}}, \bibinfo{pages}{1527--1605}
  (\bibinfo{year}{2012}).

\bibitem{Rosenfelder:1999cd}
\bibinfo{author}{Rosenfelder, R.}
\newblock \emph{\bibinfo{journal}{Phys. Lett.}}
  \textbf{\bibinfo{volume}{B479}}, \bibinfo{pages}{381--386}
  (\bibinfo{year}{2000}).

\bibitem{Sick:2003gm}
\bibinfo{author}{Sick, I.}
\newblock \emph{\bibinfo{journal}{Phys. Lett.}}
  \textbf{\bibinfo{volume}{B576}}, \bibinfo{pages}{62--67}
  (\bibinfo{year}{2003}).

\bibitem{Mohr:2008fa}
\bibinfo{author}{Mohr, P.~J.}, \bibinfo{author}{Taylor, B.~N.} \&
  \bibinfo{author}{Newell, D.~B.}
\newblock \emph{\bibinfo{journal}{Rev. Mod. Phys.}}
  \textbf{\bibinfo{volume}{80}}, \bibinfo{pages}{633--730}
  (\bibinfo{year}{2008}).

\bibitem{Belushkin:2006qa}
\bibinfo{author}{Belushkin, M.~A.}, \bibinfo{author}{Hammer, H.~W.} \&
  \bibinfo{author}{Meissner, U.~G.}
\newblock \emph{\bibinfo{journal}{Phys. Rev.}} \textbf{\bibinfo{volume}{C75}},
  \bibinfo{pages}{035202} (\bibinfo{year}{2007}).

\bibitem{Sick:2011zz}
\bibinfo{author}{Sick, I.}
\newblock \emph{\bibinfo{journal}{Few-Body Syst.}}
  \textbf{\bibinfo{volume}{50}}, \bibinfo{pages}{367--369}
  (\bibinfo{year}{2011}).

\bibitem{Zhan:2011ji}
\bibinfo{author}{Zhan, X.~{\it et al.}.}
\newblock \emph{\bibinfo{journal}{Phys. Lett.}}
  \textbf{\bibinfo{volume}{B705}}, \bibinfo{pages}{59--64}
  (\bibinfo{year}{2011}).

\bibitem{Sick:2008aa}
\bibinfo{author}{Sick, I.}
\newblock \emph{\bibinfo{journal}{Lecture Notes in Physics}}
  \textbf{\bibinfo{volume}{745}}, \bibinfo{pages}{57--77}
  (\bibinfo{year}{2008}).

\bibitem{Krauth:2016aa}
\bibinfo{author}{Krauth, J.~J.}
\newblock \bibinfo{title}{Charge radii from hydrogen-like muonic atoms -
  preliminary value for charge radius of the deuteron from muonic atom shown in
  table 1}.
\newblock \bibinfo{howpublished}{PSAS Conference, Jerusalem}
  (\bibinfo{year}{2016}).

\bibitem{VanOrden:2016aa}
\bibinfo{author}{Van~Orden, J.} (\bibinfo{year}{2016}).
\newblock \bibinfo{note}{Private communication}.

\end{thebibliography}

\end{document}